\documentclass[11pt]{article}
\usepackage{amsmath}
\usepackage{graphicx}
\def\D{\Delta}
\def\d{\delta}
\def\L{\Lambda}

\def\S{\Sigma}
\def\G{\Gamma}

\def\e{\epsilon}
\def\s{\sigma}
\def\S{\Sigma}

\def\O{\Omega}

\def\a{\alpha}

\def\m{\mu}
\def\n{\nu}

\def\s{\sigma}
\def\p{\pi}

\def\f{\phi}
\def\F{\Phi}
\def\th{\theta}
\def\t{\tau}
\def\e{\epsilon}

\def\vf{\varphi}

\def\ch{{\cal H}}
\def\cd{{\cal D}}
\def\cg{{\cal G}}

\def\pa{\partial}

\def\det{\textrm{det}}

\def\iu{\textrm{i}}
\def\R{{\bf R}}

\def\bv{{\Big |}}

\newcommand{\be}{\begin{equation}}
\newcommand{\ee}{\end{equation}}
\newcommand{\bea}{\begin{eqnarray}}
\newcommand{\eea}{\end{eqnarray}}

\begin{document}

\begin{center}
\bf{\Large Physical States and Transition Amplitudes in Piecewise Flat Quantum Gravity}
\end{center}

\bigskip
\begin{center}
{\bf Aleksandar Mikovi\'c} 
 \\
Departamento de Inform\'atica e Sistemas de Informac\~ao \\
Universidade Lus\'ofona\\
Av. do Campo Grande, 376, 1749-024 Lisboa, Portugal\\
and\\
Grupo de Fisica Matem\'atica\\
Instituto Superior T\'ecnico, Av. Rovisco Pais, 1049-001 Lisboa, Portugal
\end{center}

\centerline{E-mail: amikovic@ulusofona.pt}
\bigskip
\bigskip

\begin{abstract} 
We show how the path integral for gravity and matter on a piecewise flat spacetime can be used to define the physical quantum gravity states and the related transition amplitudes. The physical states are given by the path integrals for open manifolds from a certain topological class, while the corresponding transition amplitudes are obtained by gluing two such open manifolds into a closed one and calculating the corresponding path integral. We also solve the problem of how to associate a quantum gravity state to an effective action. This is done by using the path integral for a closed manifold from the transition-amplitude topological class so that the state which corresponds to the effective action can be choosen to be the Hartle-Hawking or the Vilenkin wavefunction for the vacuum manifold component of the transition-amplitude manifold.
\end{abstract}

\newpage
\section{Introduction}

A path integral (PI) quantization of General Relativity (GR) is the most natural approach to quantization of gravity, because of its geometric nature and conceptual simplicity, see \cite{Hk, HH, Hbb, cdt, MVb}. The most straightforward approach to the problem of constructing the GR path integral is using the Regge calculus, for a review and references see \cite{Hbb, MVb}. The two main difficulties in the quantum Regge calculus (QRC) approach are the definition of the Lorentzian path integral \cite{bdqs, tv, bd, fl, lf, rl}, and the definition of the smooth-manifold limit \cite{Hbb, cdt}. In the Casual Dynamical Triangulations (CDT) approach \cite{cdt} these problems are solved by constructing a finite Lorentzian path integral and by using the Wilsonian approach in order to find the smooth-manifold limit. This was done by replacing the integral over the edge lengths of a spacetime triangulation by a sum of $e^{iS_R/l_P^2}$ over causal triangulations, where $S_R$ is the Regge action and $l_P$ is the Planck length, while the edge lengths take only two fixed values. By restricting the number of triangulations in the sum to be finite and by performing a Wick rotation one obtains a finite path integral \cite{cdt}. However, the main drawback of the CDT approach is the absence of analytical techniques in the case of 4d spacetimes, so that the only way to obtain results is through numerical simulations, which makes the CDT approach difficult to use.

Given the difficulties of the CDT approach, it was proposed in \cite{M} that the smooth-manifold limit problem can be solved in the QRC approach by postulating that the short-distance structure of the spacetime is not a smooth manifold, but it is given by a piecewise linear (PL) manifold which is piecewise flat and which corresponds to a spacetime triangulation with a large number of the 4-simplices. The path-integral finitennes problem is then solved by introducing a non-trivial PI measure \cite{M,M1}, which is a new and efficent  technique for obtaining finiteness.

We call such a quantum gravity theory a picewise flat quantum gravity (PFQG) and the smooth spacetime is then recovered as a PL manifold where  the number of edge lengths is large in a macroscopic spacetime region, while the average edge length is microscopicaly small. In this case the corresponding effective action (EA) can be approximated by the effective action for a quantum field theory (QFT) with a cutoff, where the QFT is defined on the corresponding smooth-manifold region \cite{M, MVb}. The QFT cutoff is then determined by the average edge length \cite{MVb}, and the corresponding effective action is well defined because of the existence of a physical cutoff. In this way one recovers the correct semiclassical limit of the PFQG theory, since the semiclassical expansion of the PFQG effective action is then given by the usual perturbative QFT effective action for GR with matter. 

The PFQG theory is also well defined non-pertubatively, since the path integral is finite for a certain class of PI measures \cite{M1,M2}, so that the effective action and the correlation functions are defined non-perturbatively. 

In the standard quantum mechanics (QM) and in its direct generalization given by a flat-spacetime QFT, it is clear how to express the various physical amplitudes as path integrals, see \cite{qft}. However, when gravity is dynamical, this relationship is less clear.  The Hartle-Hawking (HH) construction \cite{HH}, as well as the Vilenkin construction \cite{V},  solve the conceptual problem of how to construct a physical wavefunction of the Universe (WFU) by using a path integral. However, due to the lack of the definition of the full QG path integral, the HH and the Wilenkin constructions have been only implemented in the context of minisuperspace models, see for example \cite{Hrev, HHH, JHH, V2}. Since the PFQG theory gives a construction of a full QG path integral \cite{M1,M2}, one can now discuss the construction of a general WFU. In this paper we will show how to construct a general WFU and also how to construct the related physical amplitudes.

The fundamental object in PFQG theory is the effective action, since the semiclassical limit is defined by using the effective action \cite{M1, MVb}. However, from the QM perspective, it is not clear what is the physical state which corresponds to the PFQG effective action. Note that in the QFT case, the effective action is defined with respect to the vacuum state, which is defined as the ground state of the Hamiltonian. However, in the QG case, such a definition cannot be used because there is no unique Hamiltonian\footnote{A Hamiltonian in quantum gravity is defined with respect to the choice of a time variable, see \cite{MVb} for a review and references.} and every QG Hamiltonian is unbounded from below. Fortunately, we have the HH and the Vilenkin constructions, and we will show how these constructions can be used to associate a state to the QG effective action.

Note that in \cite{M,M1,M2} it was observed that the PFQG effective action can be defined by using the path integral for a spacetime manifold without a boundary, or by using the same manifold, but with two spatial boundaries. The later case is dictated by the Cauchy problem for the classical action, but it remained unclear what is the relationship between these two constructions. Also, when using the manifolds with zero or two boundaries to construct the effective action, it was not clear what are the QG states which correspond to each of the QG effective actions. In this paper we will clarify these issues.

In section 2 we briefly review the PFQG path integral. In section 3 we briefly review the QFT effective action and the construction of the PFQG effective action. In section 4 we explain how to extend the basic QM path integral formulas to the PFQG theory in order to give the path integral definitions of the physical states and their transition amplitudes. In section 5 we show how to construct a state that that can be associated to the PFQG effective action. In section 6 we present our conclussions.

\section{Quantum gravity path integral}

In this section we review the basics of PFQG theory, see \cite{MVb, M2}. Let $M$ be a closed, compact and smooth 4-dimensional manifold and let $T(M)$ be a piecewise linear manifold which corresponds to a regular finite triangulation\footnote{We consider a triangulation to be regular if the dual one-simplex is given by a connected 5-valent graph with more than 5 vertices.} of $M$. We will restrict the topology of $M$ to be
\be M = M' \sqcup \left(\S\times I \right)\sqcup M'' \,, \label{pfm}\ee
where
\be \pa M' = \pa M'' = \S \,,\ee
and $I=[0,t]$, see Fig. 1.
\begin{figure}[htpb] 
\centering
\includegraphics[width=0.6\textwidth]{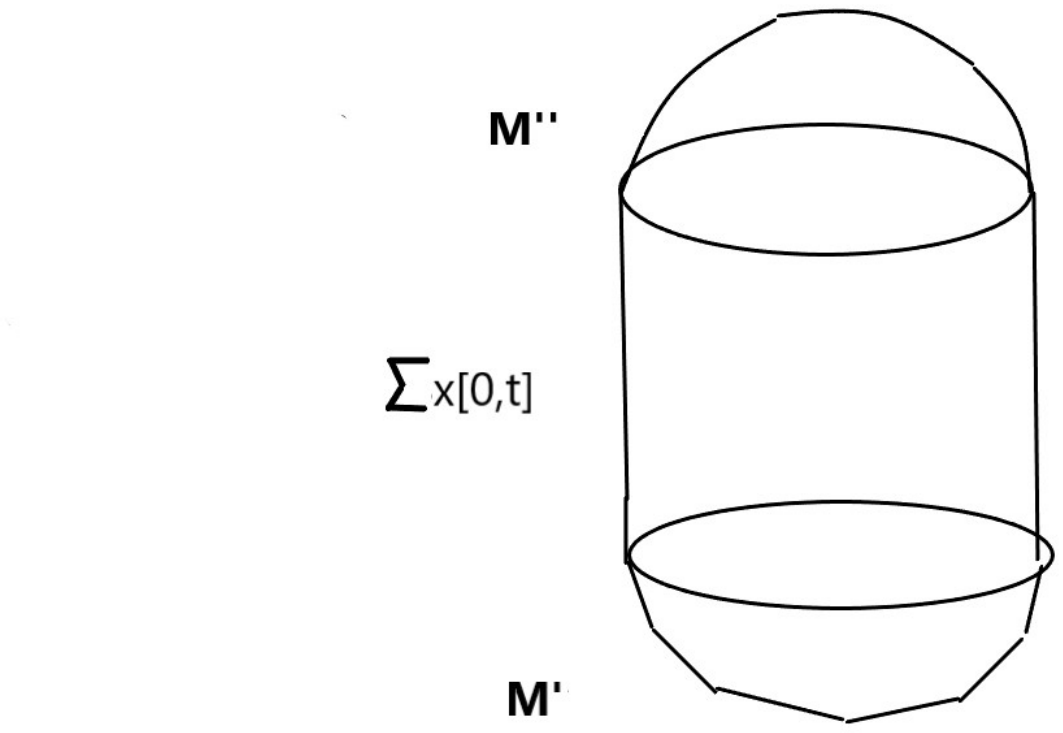}
\caption{Topology of a PFQG compact spacetime manifold}
\end{figure} 

The 3-manifold $\S$ and the 4-manifolds $M'$ and $M''$ can be of arbitrary topologies, with the restriction that the manifolds $M'$ and $M''$ should contain a submanifold which is the 3-manifold $\S$. In this paper we will be interested in the case when $M' = M''$.

Let us choose a triangulation $T(M)$, and let $\{L_\e |\, \e\in T_1(M)\}$ be a set of the edge lengths  such that  $L_\e^2 \in \R$, i.e.
$L_\e \in \R_+$ (a spacelike edge) or $L_\e \in i\R_+$ (a timelike edge). We do this in order to be able to construct Lorentzian signature metrics on $T(M)$.

A metric on $T(M)$,  which is flat in each 4-simplex $\s$ of $T(M)$, is given by 
\be G_{\m\n}(\s) = L_{0\m}^2 + L_{0\n}^2 - L_{\m\n}^2 \,,\label{cmm}\ee
where the five vertices of $\s$ are labeled as $0,1,2,3,4$ and $\m,\n = 1,2,3,4$. This metric is known as the Cayley-Menger (CM) metric. 

The CM metric is not dimensionless and hence it is not diffeomorphic to 
\be g_{\m\n}(\s) = 
\begin{pmatrix}
-1 & 0 & 0 & 0 \\
0 & 1 & 0 & 0 \\
0 & 0 & 1 & 0 \\
0 & 0 & 0 & 1 
\end{pmatrix}\,,\label{mink}\ee
see \cite{MVb}. This can be corrected by using a dimensionless PL metric 
\be g_{\m\n}(\s) = {G_{\m\n}(\s) \over |L_{0\m}||L_{0\n}|} \,.\label{eplm}\ee
The metric (\ref{eplm}) is obtained by requiring that the scale factor is polynomial and that the Minkowski metric (\ref{mink}) can be obtained by using linear coordinate transformations.

The Einstein-Hilbert (EH) action on $M$ is given by
\be S_{EH} = \int_M \sqrt{|\det g|}\, R(g) \, d^4 x  \,,\label{eh}\ee
where $R(g)$ is the scalar curvature associated to a metric $g$. On $T(M)$ the EH action becomes the Regge action 
\be S_R (L) = \sum_{\D\in T(M)} A_\D (L)\, \d_\D (L) \,,\label{era}\ee
when the edge lengths correspond to a Euclidean PL geometry. $A_\D$ is the area of a triangle $\D$, while  the deficit angle $\d_\D$ is given by
\be \d_\D = 2\pi - \sum_{\s \supset \D} \th_\D^{(\s)} \,,\ee
where a dihedral angle $\th_\D^{(\s)}$ is defined as the angle between the 4-vector normals associated to the two tetrahedrons that share the triangle $\D$. 

When $M$ has a boundary, one should add to $S_{R}$ the corresponding boundary term, see \cite{hs}. 

In the case of a Lorentzian geometry, a dihedral angle can be a complex number, because when two vectors belong to a timelike plane then the angle between them is complex. This follows from the Lorentzian angle definition
\be \cos\th = {\vec u \cdot \vec v \over ||\vec u||\,||\vec v||}\,,\quad \sin\th = \sqrt{1-\cos^2 \th}\,, \label{lang}\ee
and the fact that a vector from a  timelike plane can be represented as a multiple of one of the following vectors
\be \{(\pm\cosh a, \pm \sinh a)\,, (\pm\sinh a, \pm \cosh a)\} \,,\quad a\in \R \,, \ee
where 
\be \vec u \cdot \vec v = -u_0 v_0 + u_1 v_1 \,, \quad ||\vec u|| = \sqrt{-u_0^2 + u_1^2} \,,\ee
see \cite{cdt, MVb}. 

From the dihedral angle formula
\be \sin \th_\D^{(\s)} = \frac{4}{3} {v_\D v_\s \over v_\t v_{\t'}} \,, \ee
where $v_s$ is the volume $V_s$ of the simplex $s$ ($V_s \ge 0$ by definition) if the CM determinant is positive, while $v_s = \iu V_s$  if the CM determinant is negative, and by using (\ref{lang}), we can see that $\th_\D^{(\s)}$ is a complex number for a spacelike triangle, while it is real for a timelike triangle. 

Hence $\d_\D$ will be real for a timelike triangle, while
\be  \d_\D =  m_\D \frac{\p}{2} + \iu b_\D \,,\ee
for a spacelike triangle, where $m_\D \in \bf{Z}$ and $b_\D \in \R$. One can then modify the Regge action as
\be S_R(L) = Re\left(\sum_{\D'} A_{\D'} \, \frac{1}{\iu}\,\d_{\D'} \right) + \sum_{\D''} A_{\D''} \,\d_{\D''}  \,,\label{lra} \ee
where $\D'$ denotes a spacelike triangle, while $\D''$ denotes a timelike triangle. Consequently $S_R(L)$ is a real number, and it corresponds to the Einstein-Hilbert action on $T(M)$ \cite{cdt, MVb}.

Therefore the PFQG path integral is defined as
\be Z(T(M)) = \int_D \prod_{\e=1}^N dL_{\e} \, \m(L) \, e^{i S_R(L)/l_P^2} \,, \label{grpi}\ee
where $dL_\e = d|L_\e|$, $l_P$ is the Planck length and $\m(L)$ is a measure that ensures the convergence and one also requires it gives the effective action with a correct semiclassical expansion, see \cite{M,M1}. The integration region $D$ is a subset of ${\bf R}_+^N$, consistent with a choice of spacelike and timelike edges.

One can then show that the path integral (\ref{grpi}) is convergent for the measure
\be \m(L) = e^{-V_4(L)/L_0^4}\prod_{\e=1}^N \left(1+ {|L_\e|^2 \over l_0^2} \right)^{-p} \,,\label{plm}\ee
while the corresponding effective action has a correct semiclassical limit, where $V_4(L)$ is the 4-volume of $T(M)$, $L_0$ and $l_0$ are free parameters and the parameter $p$ has to satisfy $p > 1/2$, see \cite{M1}. The bound $p>1/2$ can be easily derived from the requirement of the absolute convergence, since
\be |Z| \le \int_D \prod_{\e=1}^N dL_{\e} \, \m(L) <  \prod_{\e=1}^N \int_0^\infty dL_{\e} \, \left(1+ {|L_\e|^2 \over l_0^2} \right)^{-p} \,. \ee

Note that the convergence can be also obtained without the $e^{-V_4/L_0^4}$ factor in the measure, but the exponential factor is necessary in order to obtain the correct classical limit of the effective action, see \cite{M, M1}.

When $\S$ is a non-compact manifold, we maintain a finite number of DOF by allowing non-zero $L$ vectors, i.e. $(L_1, L_2, \cdots, L_N) \ne (0, 0, \cdots, 0)$, only for a triangulation of $B_3 \times I$, where $B_3$ is  a 3-ball in $\S$.  This subset of $\S\times I$ is then glued to two 4-balls $B_4'$ and $B_4''$ in $M'$ and $M''$, see Fig. 2
\begin{figure}[htpb] 
\centering
\includegraphics[width=0.6\textwidth]{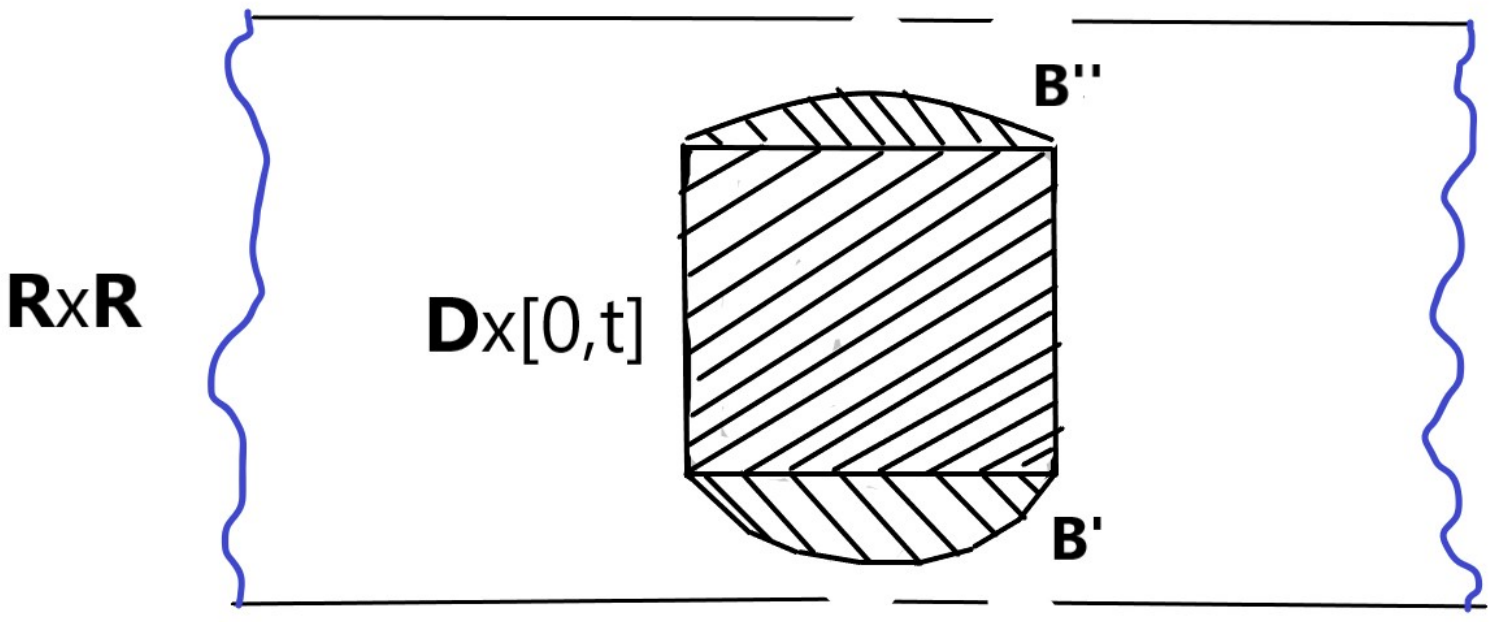}
\caption{Topology of a PFQG non-compact spacetime manifold}
\end{figure} 
Therefore one has non-zero $L$ vectors in a triangulation of a compact region of $M'\sqcup (\S\times I)\sqcup M''$ given by $B_4'\sqcup (B_3 \times I)\sqcup B_4''$.

Note that in the 2d representation of $M$ in Fig. 2,  the 3-ball $B$ is represented by an interval $D$, while the 4-balls $B'$ and $B''$ in $M'$ and $M''$ are represented by half-discs.

We can include the matter DOF by giving the matter fields at the vertices of the triangulation \cite{M2}. Let $\f = \{\f_\a \,| \,\a = 1,2, ...,c\}$ be a set of $c$ matter field components at a spacetime point. The corresponding gravity plus matter path integral will be given by
\be Z = \int_D d^N L \,\mu(L)\, e^{i S_R(L)/l_P^2}\, Z_m(L) \,,\label{gsfpi}\ee
where
\be Z_m (L) = \int_{D_m} \prod_{\a=1}^c d^n \f_\a \, e^{i S_m(\f, L)/\hbar} \,,\ee
is the matter path integral. Here $S_m(\f,L)$ denotes the matter action on $T(M)$, $n$ is the number of vertices in $T(M)$ and $D_m$ is the range of the matter fields components. We will consider the matter action to be that of the Standard Model, so that we will include the fermionic fields in the path integral. Hence the integration set  $D_m$ will be given by
\be \R^{c'n}\times \prod_{v=1}^n \L_{c''}(v) \,,\ee 
where $c'$ is the number of bosonic fields components, $\L_{c''}(v)$ is a $c''$-dimensional Grassman algebra at a vertex $v$ and $c' +c'' = c$.

Since the convergence of $Z_m$ is not guaranteed, we pass to a Euclidean geometry defined by the edge lengths
$$ \tilde L_\e = |L_\e| \,, $$
so that all the Euclidean edge lengths are positive real numbers. This is equvalent to a Wick rotation where $\tilde L_\e = L_\e$ if $\e$ is a spacelike edge and $\tilde L_\e = (-\iu)L_\e$, if $\e$ is a timelike edge.

By intregating the fermionic fields we obtain
\be \tilde Z_m ( \tilde L) = \int_{D'_m}\prod_{\a'} d^n  \f'_{\a'} \, P(\f') e^{-\tilde S_m(\f',  \tilde L)/\hbar} \,,\label{zm}\ee
where $\tilde S_m$ is the Euclidian bosonic matter action, while $P(\f')$ is a polynomial function corresponding to the product of the fermionic and the Yang-Mills ghost fields determinants. Since $\tilde S_m (\f', \tilde L)$ is a positive polynomial function of the bosonic fields $\f'$
then the integral $\tilde Z_m$ will be convergent. Hence we will define the Lorentzian matter path integral as
\be Z_m(L) = \tilde Z_m (\tilde L)\bv_{\tilde L = w(L)} \,,\ee
where $w$ is the Wick rotation.

One can then show that the PFQG path integral (\ref{gsfpi}) is absolutely convergent for
\be p > \textrm{52,5} \,,\label{pv}\ee
see \cite{M2}. This is done by making a change of the integration variables $(L_1, L_2, \cdots, L_N)$ to $N$-dimensional spherical coordinates $(r, \th_1, \th_2, \cdots, \th_{N-1}) $, so that
\be |Z| < \int_D d^N L \,\m(L) |Z_m(L)| < \int d^N L \, \m(L) r^{\tilde c n} F_n(\th) \,,\ee
or
\be |Z| < \int_0^\infty r^{N-1 +c^*n}dr \int_\O J_N(\th) \tilde\m(r,\th) F_n(\th)d^{N-1}\th \,,\ee
where $\tilde c = 260$, $\m(L) = \tilde\m(r,\th)$ and $J_N(\th)d^{N-1}\th$ is the angular volume element.

By using the asymptotic properties of the PI measure $\tilde\m(r,\th)$ for small and large $r$, we obtain
\be |Z| < C_1 \int_0^R r^{\tilde c n + N-1} dr + C_2 \int_R^\infty r^{\tilde c n + N-1 -2pN} dr \,,\ee
so that we can guarantee the absolute convergence of the PFQG path integral if 
\be \tilde c n + N(1-2p) < 0 \,,\ee
or
\be {\tilde c\over 2p-1} < {N\over n} \,.\label{rtb}\ee

For a regular triangulation we have $N/ n \ge 5/2$, so that if 
\be \frac{\tilde c}{2p-1} < \frac{5}{2} \,,\ee
 then (\ref{rtb}) will be satisfied, and we obtain for the Standard Model the bound (\ref{pv}).

Hence for the values of the parameter $p$ given by (\ref{pv}), we have a finite path integral, and this makes it possible to construct the PFQG states, the transition amplitudes and the effective action.

\section{Quantum gravity effective action}

The idea of the effective action was introduced in the context of a QFT as the gnerating functional of the one-particle irreducible correlation functions 
\be \G[\f(x)] = \sum_{n=2}^\infty \int d^{4n} x \, \G_n(x_1, ... , x_n) \f(x_1)\cdots \f(x_n) \,,\ee
see \cite{qft}.

Note that
\be\G_n (x_1, \cdots , x_n) = \langle 0|\hat\f(x_1) \cdots \hat\f(x_n) |0\rangle\bv_{1PI}\,,\label{eacf}\ee
where $\langle ...\rangle|_{1PI}$ means to take only the one particle irreducible graphs in the Feynman diagram expansion of the $n$-point correlation function. Also note that
\be \langle 0|\hat\f(x_1) \cdots \hat\f(x_n) |0\rangle = \int \cd\f \,\f(x_1) \cdots \f(x_n) e^{iS(\f)/\hbar}\,,\label{cf}\ee
where $S(\f)$ is the classical action. Hence the formulas (\ref{eacf}) and (\ref{cf}) establish a clear relationship between the effective action, the path integral and the QFT vacuum state $|0\rangle$.

It is not difficult to show that the effective action satisfies the equation
\be e^{i\G[\f]/\hbar} = \int \cd\f' \exp\left(\frac{i}{\hbar}S[\f + \f'] - \frac{i}{\hbar}\int_M \frac{\d\G}{\d\f(x)}\,\f'(x) \,d^4 x\right)\,, \ee
see \cite{M1, eae}. This equation gives a nonperturbative definition of the QFT effective action, but also it can be directly applied to the PFQG case, where it becomes an integro-differential equation, see \cite{M,M1,MVb}.

More precisely, a non-perturbative effective action $\G(L,\F)$ on $T(M)$, where $\F =(\f_1,\f_2,...,\f_n)$, is determined by the equation
\bea  e^{i\Gamma(L,\F)/\hbar} &=& \int_{D(L)} d^N  l \int_{D_m}  d^{cn} \vf \, \mu (L +l)\, e^{iS (L+l, \F + \vf)/\hbar - i\sum_{\epsilon} \G'_\epsilon (L,\F)l_\epsilon /\hbar}\cr
&\,& \qquad\qquad\qquad\qquad\quad  e^{ -  i\sum_{\p} {\Gamma}'_\p (L,\F) \vf_\p /\hbar } \,,\label{eagm}\eea
where $D(L)$ is the PI integration region $D$ translated by a vector $-L$, see \cite{MVb, M}, while $c$ is the number of components of the matter fields. Note that for the Standard Model 
\be c = c_f  + c_b = 96 + 52 = 148\,,\ee 
where $c_f$ is the number of fermionic fields componets and $c_b$ is the number of bosonic fields components, $S= S_R/G_N + S_m$ and $G_N$ is the Newton constant \cite{M2}. 

The equation (\ref{eagm}) will be well defined if the gravity plus matter path integral is finite, which is the case for $p >$ 52,5 \cite{M2}. This is a consequence of
\be |\tilde Z_m(\tilde L, J)| \le \tilde Z_m(\tilde L) \,,\ee
where
\be \tilde Z_m(\tilde L, J) = \int_{D_m}\prod_\a d^n  \f_\a \, e^{[-\tilde S_m(\f,  \tilde L) + i J\f ] /\hbar} \,,\ee
is the matter path integral with sources. This is the discretized version of the matter QFT generating functional.

The effective action defined by the equation (\ref{eagm}) has an important property: when the number of edges $N$ is large and the average edge length $\bar L$ in $T(M)$ is microscopically small, then $\G(L,\F)$ can be approximated by a QFT effective action on $M$ where the QFT is given by the momentum cut-off quantization of GR coupled to the Standard Model, while the momentum cutoff is given by $2\pi\hbar/\bar L$, see \cite{MVb}. This is a consequence of the fact that a PL function on a finite interval $I$ which is subdivided into $n$ smaller intervals of length $\bar L$, can be approximated for large $n$ by the Fourier integral of a smooth function on $I$ with the wavevector cutoff $2\pi/\bar L$.

In particular, when $|L_\e| > l_P$ and $L_0 > l_P$, the effective action $\G(L,\F)$ has a perturbative expansion
\be \G = S + \hbar\G_1 + \hbar^2 \G_2 + \cdots = { S_R \over G_N} + S_m  + \hbar\G_1 + \hbar^2 \G_2 + \cdots \,,\label{pea}\ee
see \cite{M,MVb}. This can be understood from the fact that the perturbative expansion can be rewritten as
\be {\G\over\hbar} = { S_R + \tilde S_m \over l_P^2} + \G_1 + l_P^2 \tilde\G_2 + l_P^4 \tilde\G_3 + \cdots \,,\ee
where $\tilde S_m = G_N S_m$ and $\tilde\G_k =\G_k /(G_N)^{k-1}$, $k=1,2,...$ . 

When $N$ is large and $\bar L$ is microscopically small, then one can show that the perturbative solution (\ref{pea}) of the EA equation (\ref{eagm}) is very well approximated by the classical GR action plus the classical SM action plus the corresponding QFT loop corrections \cite{M,M1}. This follows from the smooth-manifold approximation, which is defined by $N\to\infty$ and $|L_\e| = O(1/N)$ such that
\be g_{\m\n}(x) \approx g_{\m\n}(\s) \,, \quad \f_\a (x) \approx \f_\a (v)  \quad  \textrm{for}\,\, x\in\s  \,,\ee\
where $v$ is a dual vertex in $\s$, $g_{\m\n}(x)$ is a smooth metric on $M$ and $\f_\a (x)$ is a smooth matter field on $M$. Then
\be \G(L,\F) \approx \G_K [g_{\m\n}(x), \f(x)] \,,\quad x\in M\,,\ee
where $\G_K$ is the QFT effective action for the cutoff $K = 2\pi/\bar L$ and $\bar L$ is the average edge length in $M$. For $|L_\e| > l_P$ we have
\be\G (L,\F) \approx \frac{1}{G_N} S_{EH} (g) + S_m (g,\f) +  \hbar\, \G_K^{(1)}(g,\f) + \hbar^2 \,\G_K^{(2)}(g,\f) + \cdots \,,\label{smpea} \ee
where $\G_K^{(n)}$ is the $n$-loop QFT effective action for GR coupled to matter, see \cite{MV}. 

Note that validity of the perturbative QFT expansion (\ref{smpea}) requires that $|L_\e| > l_P$ so that $\bar L > l_P$. This still allows for $\bar L$ to be microscopically small. For example, the distance probed in the LHC experiments is of the order of $10^{-20}$m, while $l_P = O(10^{-34}\textrm{m})$. When $\bar L \le l_P$, one can still use the smooth-manifold approximation, but the corresponding QFT effective action will not be given by the perturbative QFT expansion.

One can also add the cosmological constant (CC) term to the Regge action, so that
\be S_R(L) \to  S_R(L) + \L_c V_4 (L) \, . \ee
Then the condition for the semi-classical expansion of the effective action, $|L_\e| \gg l_P$ and $L_0 \gg l_P$, is substituted by 
\be |L_\e| \gg l_P \,,\quad  L_0 \gg \sqrt{l_P L_c} \,,\ee
where $|\L_c| = 1/L_c^2$ is a free parameter \cite{MV}. 

Hence the PFQG theory for the Standard Model defined by the path integral (\ref{gsfpi}), with the measure (\ref{plm}), is finite for $p>52,5$ and such a theory has the correct classical limit, which is GR coupled to the Standard Model on $T(M)$. When $N$ is large, while $\bar L$ is microscopically small in a macroscopic spacetime region, one recovers GR coupled to the Standard Model on a smooth spacetime.

Since the equation (\ref{eagm}) for the effective action for gravity coupled to matter is well defined when the PFQG path integral is finite, we can then determine the non-perturbative effective action. Note that the EA equation (\ref{eagm}) by itself is not sufficient to determine the QG analog of the QFT vacuum state. We will address this problem in the next section.

\section{Path integral and physical states}

In order to find a relationship between the path integral and a state in quantum gravity, we start from the path-integral representation of the propagator in quantum mechanics
\be  G (q_f, q_i;t) =\langle q_f | \hat U (t) | q_i \rangle = \left[\int \cd q \, e^{i\int_0^t d\t  L(\dot q, q)/\hbar}\right]_{q(0) = q_i}^{q(t) = q_f}\,,\label{piprop} \ee
where $q$ denotes the set of generalized coordinates of the system, $L(\dot q, q)$ is the Lagrangian, $q_i$ denotes the set of initial $q$ values at time $t=0$, while $q_f$ denotes the set of final values at a time instant $t$ and $\hat U(t)$ is the evolution operator. 

In the QG case, when  $M=\S\times I$, then the path integral given by $Z(T(M))$ corresponds to the formula (\ref{piprop}), since
\be G (l_f,\f_f ; l_i,\f_i ) = \langle l_f,\f_f | \hat U (t) | l_i,\f_i \rangle = \left[Z (T(\S\times I))\right]_{L|_i = l_i, \,\F |_i = \f_i}^{ L|_f = l_f , \,\F|_f = \f_f } \,,\label{qgprop}\ee
where the labels $l$ and $\vf$ indicate the vectors $L$ and $\F$ on the initial and the final triangulation of the spatial hypersurface $\S$. The label $t$ is defined by an embedding of $T(\S\times I)$ in $\R^5$ such that the distance from $\S_i$ and $\S_f$ is $t$, i.e. $I = [0,t]$, see Fig. 3. 
\begin{figure}[htpb] 
\centering
\includegraphics[width=0.6\textwidth]{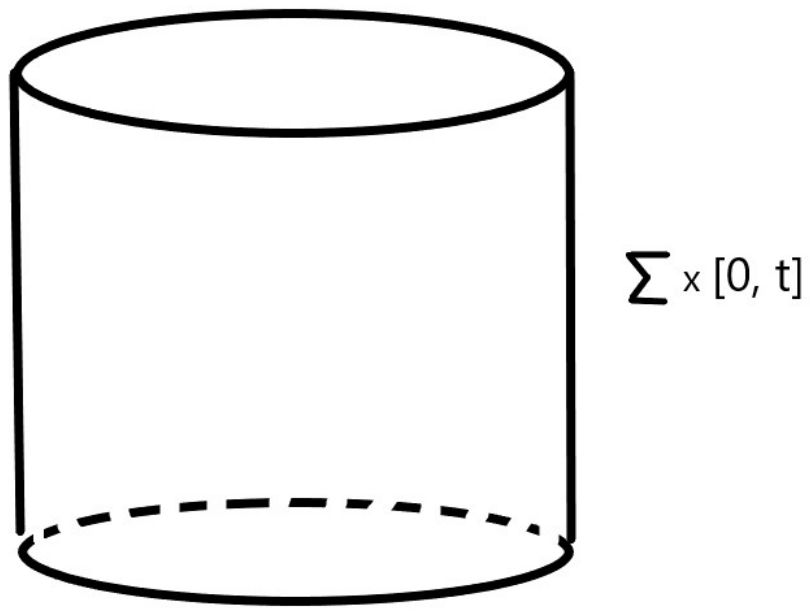}
\caption{Topology of the QM propagator manifold.}
\end{figure} 

More precisely, the described embedding of $T(\S\times I)$ in $\R^5$ fixes the time-reparametrization symmetry, since the distance from the initial $T(\S)$ to the final $T(\S)$ in $\R^5$ is fixed to be $t$. In this way one fixes the residual time-reparametrization symmetry of the reduced phase space (RPS) Schrodinger equation
\be \left( i\hbar {\pa\over \pa \t} - \hat H (p,q, \t)\right)\Psi(q,\t) = 0 \,,\label{rpse}\ee
where $H$ is the reduced phase space Hamiltonian, $(p, q)$ is a point of the reduced phase space and $\t$ is the corresponding time variable. 

The Schrodinger equation (\ref{rpse})  is a consequence of the time-variable gauge fixing 
\be T(x,\t) = \t \,,\ee
where $(x,\t)$ are coordinates of a point of $\S\times I$. The time variable $T(x,\t)$ is defined by a canonical transformation 
\be (\p^{kl},h_{kl}) \to (P,T, P_k, X_k, \tilde p, \tilde q ) \,,\label{tvct}\ee
where $(\p^{kl},h_{kl})$, $k,l = 1,2,3$, are the Arnowit-Deser-Misner variables, while $(\tilde p, \tilde q)$ are the phase-space variables of the spatial metric independent DOF. 

The canonical transformation (\ref{tvct}) must be such that the diffeomorphism constraints $\cg_k (\p,h,p',q')$ and the Hamiltonian constraint $\cg(\p,h,p',q')$, where $(p',q')$ are the matter phase-space variables,  transform  into
\be {\tilde\cg}_k = P_k - f_k (p,q, X, T) = 0 \,,\quad \tilde\cg = P - \vf (p,q, X, T) = 0 \,, \label{rpsc}\ee
where $(p,q) =(\tilde p,p', \tilde q,q')$.

Consequently we have
\be S_{ADM} + S_m = \int_0^t d\t  \int_\S d^3 x \left( P\dot T + P^k\dot X_k + p\dot q - n^k {\tilde\cg}_k - N \tilde\cg \right)\,. \ee
By choosing the gauge $X_k = x_k$ and $T=\t$ we fix the 3-diffeomorphism and the time-reparametrization summetry and solve the constraints (\ref{rpsc}) for $P$ and $P_k$, so that the dynamics of the independent variables is described by the RPS action 
\be  \tilde S_{ADM} + S'_m = \int_0^t d\t  \int_\S d^3 x \left( p\dot q - \tilde\ch (p,q, x,\t) \right)\,,\ee
where $\tilde\ch = - \vf (p,q, x, \t)$, see \cite{MVb}. Hence the RPS Hamiltonian is given by
\be H = \int_\S d^3 x\, \tilde\ch (p,q, x,\t) \,, \ee
and after the quantization one obtains the Schrodinger equation (\ref{rpse}).

When passing to a PL manifold $T(\S\times I)$, the spatial coordinate label $x$ becomes a discrete variable associated with the vertices of the triangulation, while the parameter $\t$ is associated to the distance between the inital and the final spatial section. We also assume that the triangulation $T(\S\times I)$ is causal, i.e. consists of $n$ triangulations $T_k (\S)$, $k=0,1,2,...,n$, with spacelike edge lengths, such that $T_{k}$ and $T_{k+1}$ are connected with timelike edge lengths $t_\e^{(k)}$, $k = 0,1,2,...,n-1$. The fixing of distance between $T_0$ and $T_n$ requires a constraint 
\be t_{\e}^{(0)} +t_{\e'}^{(1)} + ... = t\,, \label{tgf}\ee 
for a set of $n$ timelike edges forming a timelike line from a point in $T_0$ to a point in $T_n$.

The Schrodinger equation (\ref{rpse}) still has the residual time repara- metrization symmetry given by $\t\to\t' = f(\t)$. We fix it by setting the distance between the inital and the final spatial hypersurface in $\R^5$ to be $t$, see the equation (\ref{tgf}), so that $\t =t$. Hence $(l,\f) \cong (q,t)$, so that is the reason we did not include the label $t$ for $G (l_f,\f_f ; l_i,\f_i )$ in the equation (\ref{qgprop}). 

The QM propagator (\ref{piprop}) does not say anything about the evolution of an arbitrary state, so that we need to consider the time-evolution formula
\be |\Psi(t)\rangle = \hat U(t) |\Psi(0)\rangle \label{ste}\,,\ee
where $|\Psi(0)\rangle$ is the initial state of the system, while the evolution operator of the system $\hat U(t)$ obeys the Schrodinger equation
\be i\hbar{d\hat U\over dt} = \hat H \hat U(t) \,, \label{scheq}\ee
where $\hat H$ is the Hamiltonian operator for the quantum system.

Hence the state of the system obeys the usual Schrodinger equation 
\be i\hbar{\pa\Psi(q,t)\over \pa t} = \langle q|\hat H | \Psi(t)\rangle = \hat H_q \Psi(q,t) \,, \label{scheq} \ee
where $\hat H_q$ is the Hamiltonian operator in the $q$-representation. Therefore the equation (\ref{ste}) describes the time evolution of a state from an initial state $|\Psi(0)\rangle$, which can be freely chosen. In the QG framework, the Schrodinger equation (\ref{scheq}) corresponds to the reduced phase space Schrodinger equation (\ref{rpse}). 

Hence the path integral $Z(T(M))$ for
\be M = M' \sqcup \left(\S\times I \right)\,, \label{tevolm}\ee
see Fig. 4,
\begin{figure}[htpb] 
\centering
\includegraphics[width=0.6\textwidth]{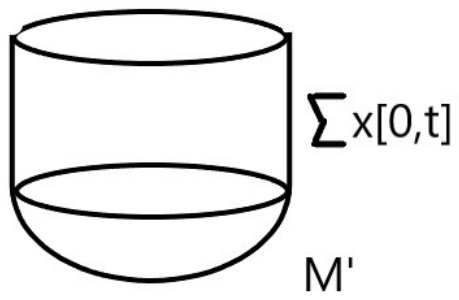}
\caption{Topology of the time-evolution manifold.}
\end{figure} 
corresponds to a wavefunction
\be \Psi(q,t) = \langle q|\Psi(t)\rangle = \langle q|\hat U(t) |\Psi(0)\rangle \label{qste}\,,\ee
where $(q, t) \cong (L|_{T(\S)}  ,\, \F|_{T(\S)} ) = (l, \f)$, so that
\be\Psi (l,\f) = \left[Z (T(M))\right]_{L|_{T(\S)}  = l ,\, \F|_{T(\S)}  = \f} \,,\label{pifst}\ee
while $I=[0,t]$. 

The state-sum formula (\ref{pifst}) for $\Psi(l,\f)$ on $\S$ is the PFQG analog of the Hartle-Hawking wavefunction \cite{HH}. Note that in the usual interpretation of the HH wavefunction, one takes $t\to 0$, so that $M=M'$, see Fig. 5,
\begin{figure}[htpb] 
\centering
\includegraphics[width=0.6\textwidth]{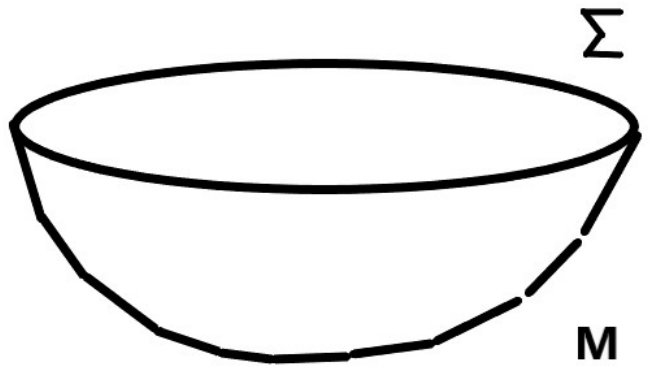}
\caption{Topology of the Hartle-Hawking manifold.}
\end{figure} 
so that $\Psi(l,\f)$ is identified with the initial wavefunction of the Universe (WFU).

Also note that in the $t=0$ case of Fig. 4, which is given by Fig. 5, one usually takes the Euclidean path integral, which seems odd from the path integral perspective, since the full path integral for QG should be defined in the Lorentzian signature (as is the case in PFQG theory). Hence, if one constructs the HH state in the case of Fig. 5 by using the Euclidean path integral, the corresponding $\Psi_0$ may not be well defined because the factor $e^{-S(L,\F)/\hbar}$ in the full Euclidean path integral is not bounded. One may try to remedy this problem by using the $e^{iS(L,\F)/\hbar}$ factor, but then the corresponding wavefunction
may not satisfy the Wheeler-de Witt (WdW) equation. However, from the point of view of Quantum Mechanics, this is not a problem, since $\Psi_0(l,\f)$ can be an arbitrary wavefunction. But in order to obtain a physical evolution, one should use the Lorentzian path integral in the $\S\times[0,t]$ component of the spacetime of Fig. 4. Hence the wavefunction of the Universe (\ref{pifst}) will certainly satisfy the WdW equation for $t > 0$. More precisely, it will satisfy the PL version of the WdW equation on $T(\S)$, see \cite{htw, ww, bf}. This is a consequence of the fact that the wavefunction (\ref{pifst}) satisfies the RPS Schrodinger equation, which is obtained by classically solving the Hamiltonian constraint, see equation (\ref{rpsc}).

However, if one uses the Lorentzian path integral in $T(M')$ in order to construct $\Psi_0 (l,\f)$, then such a $\Psi_0$ will definitely obey the PL version of the WdW equation provided that $M'$ has a topology of a 4-cone with a boundary given by the 3-manifold $\S$ \cite{MVb, M3}. This type of quantum state is known as the Vilenkin wavefunction \cite{V} and it is also called the tunneling wavefunction.

\section{Effective action and quantum gravity states}

In order to find the relationship between the effective action and a state in PFQG theory, we will now consider the following QM formula
\be \langle \Psi_2 | \hat U (t) |\Psi_1 \rangle = \int_{D'} d^{m} q_f \int_{D'} d^{m} q_i \Psi_2^*(q_f) G (q_f, q_i, t) \Psi_1 (q_i) \,,\label{tamplitude}\ee
where $\Psi_1$ and $\Psi_2$ are the initial and the final wavefunction, while $D'\subseteq \R^{m}$, is the domain of the generalized coordinates.

Let us then consider the PFQG path integral for the following smooth manifold
\be M = M' \sqcup \left(\S\times I \right)\sqcup M'\,,\label{evtop} \ee
see Fig. 6. 
\begin{figure}[htpb] 
\centering
\includegraphics[width=0.6\textwidth]{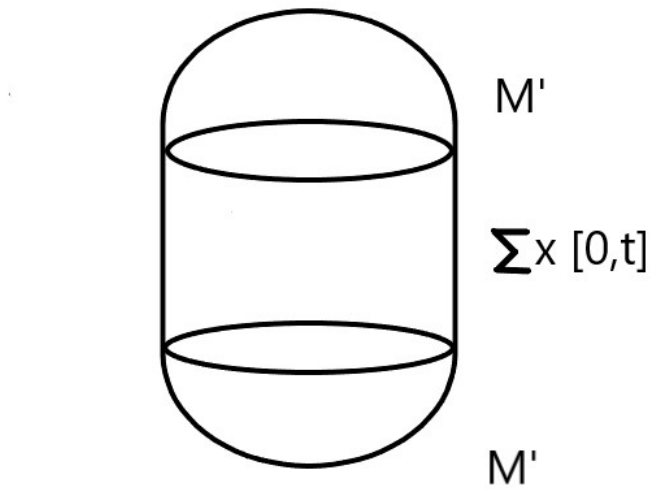}
\caption{Topology of the spacetime manifold for the expectation value of the evolution operator.}
\end{figure} 

By comparing (\ref{tamplitude}) to $Z(T(M))$ we conclude that  $\Psi_1 = \Psi_2 = \Psi_0$ and $I=[0,t]$, so that
\be Z(T(M)) = \langle\Psi_0|\hat U(t) |\Psi_0 \rangle \,.\ee

Note that if $\Psi_0$ is obtained by using a Euclidean path integral, it will be real. However, since the convergence of the Euclidean path integral is not always guaranteed, because the factor $e^{-S_E(L,\F)/\hbar}$ in the integrand can be arbitrarily large, one could then replace this factor by the factor $e^{iS_E(L,\F)/\hbar}$, where $S_E$ is the Euclidean action, or one can use the Lorentzian PI factor $e^{iS(L,\F)/\hbar}$. In any of these cases the wavefunction wil not be real, so that in order to include the $\Psi_2^*$ factor from the formula (\ref{tamplitude}),  the corresponding path integral should be calculated by using the $e^{-iS_E(L,\F)/\hbar}$ or $e^{-iS(L,\F)/\hbar}$ factor.

Let us now consider the manifold topology  (\ref{evtop}) with $I=[-t,t]$, see Fig. 7. 
\begin{figure}[htpb] 
\centering
\includegraphics[width=0.6\textwidth]{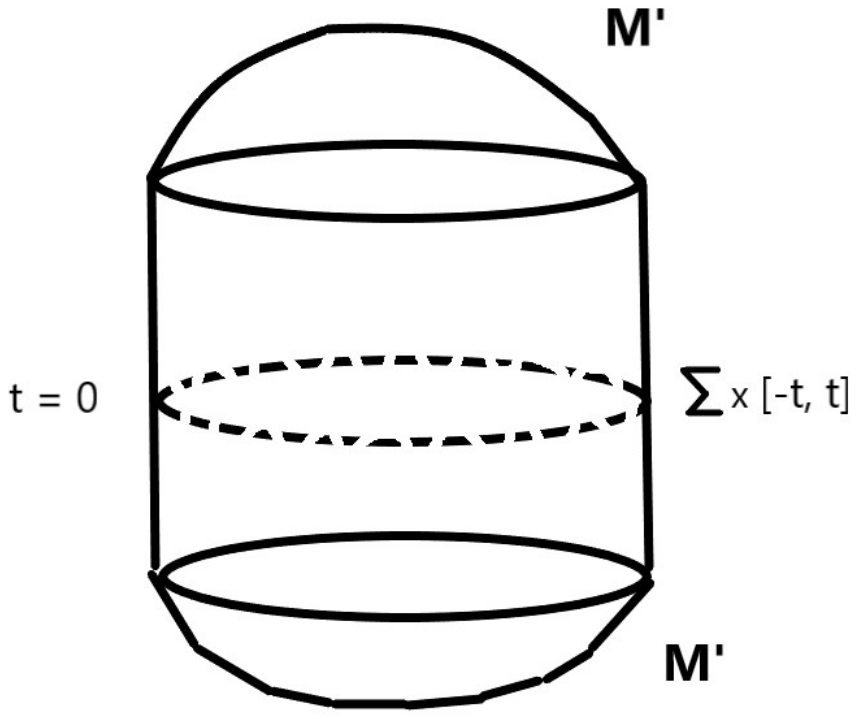}
\caption{Topology of the effective action manifold}
\end{figure} 
This case is relevant for the Heisenberg operator expectation values, because of the QM formula
\be q^{(H)}_\a (t) =\langle \Psi_0 | \hat U(-t) \, \hat q_\a (0) \, \hat U (t) |\Psi_0 \rangle \,,\label{hev} \ee
where $q_\a$ is a component of the vector $q$.

Note that there is also the following relation
\be q^{(H)}_\a (t)  = -\frac{i}{\hbar}\left[{\pa\over \pa J_\a (0) } Z_J (T(M))\right]_{J = 0} \,,\label{qgev}\ee
where $Z_J$ is the state sum $Z$ with a source term.  By comparing the relations (\ref{hev}) and (\ref{qgev}) and by taking into account the EA equation (\ref{eagm}), we conclude that the effective action defined by a manifold of Fig. 7, will correspond to the state $\Psi_0$ which corresponds to the HH state for the vacuum manifold $M'$.

In order to have an exact correspondence with the relation (\ref{hev}), the path integral $Z(T(M))$ has to be evaluated by using the action 
\be \tilde S(L,\F) = - S(L,\F)\,,\ee
in the $\S\times [0,t]$ region, while in the $\S\times [-t,0]$ region we use the plus sign. 

\section{Conclusions}

A physical state of the PFQG theory is given by the path integral for an open manifold of the topology (\ref{tevolm}), see Fig. 4. The corresponding wavefunction can be explicitely calculated, given the triangulations $T(M')$ and $T(\S\times [0,t])$, since the PFQG path integral (\ref{gsfpi}) is finite. One can also calculate the corresponding transition amplitudes by using the path integrals for the closed manifolds of Fig. 6 and Fig. 7.

One can also calculate the full Hartle-Hawking wavefunction, provided that one fixes the spatial manifold $\S$ and the 4-manifold $M'$, as well as the corresponding triangulations $T(\S)$ and $T(M')$. Note that the explicit expressions for the HH wavefunctions can be obtained  in the case of simple triangulations which are the analogs of the minisuperspace models, see \cite{MVb, M3}.

Also note that for a fixed spatial hypersurface $\S$ one can have many different HH states, since each such HH state depends on the choice of the vacuum manifold $M'$ and its triangulation. If we define a HH state as the path integral for the $t\to 0$ manifold topology of Fig. 6, which is the manifold topology of Fig. 5, then the corresponding HH state does not have to obey the WdW equation, because the HH state in this case is the analog of the inital state in the QM time-evolution relation (\ref{ste}) and the initial state $|\Psi_0\rangle$ does not have to obey the Schrodinger equation. However, for $t>0$, the state $|\Psi(t)\rangle$ obeys the Schrodinger equation, so that the corresponding HH wavefunction should obey the WdW equation. More precisely, the wavefunction (\ref{pifst})  will obey the PL version of the WdW equation on $T(\S)$.

Therefore there is no unique HH state in PFQG theory, and the question of how to choose a HH state boils down to a choice of the manifolds $M'$ and $\S$ and their triangulations. The question of whether the HH wavefunction $\Psi_0$ for the vacuum manifold $M'$ obeys the WdW equation, remains open, although one can certainly enforce the WdW equation on $\Psi_0$ by using the Lorentzian path integral and the conical topology for the vacuum manifold, i.e. by constructing the Vilenkin wavefunction, see \cite{MVb, M3} in the context of minisuperspace triangulations. 

The problem of how to associate a QG state to the effective action defined as a solution of the EA equation (\ref{eagm}) is solved by using a closed manifold $M$ of Fig. 7, so that the corresponding state is given by the HH/Vilenkin wavefunction for the vacuum manifold $M'$. This also resolves the dilemma of choosing a closed or open manifold when constructing the effective action, which was mentioned in the introduction. Hence we should always choose a closed manifold of Fig. 7.  However, the corresponding EA equations of motion describe the quantum trajectories in the $\S\times[0,t]$ component of that manifold.

Note that the formula (\ref{qgev}) suggests that $q_\a^{(H)} (t)$ can be considered as a particular solution (or a very good approximation, see \cite{cdmp}) of the EA equations of motion in the $\S\times[0,t]$ component the spacetime associated with a manifold of Fig. 7. Hence the path integral  formula (\ref{qgev}) provides a way to obtain the non-perturbative solutions of the EA equations of motion. These solutions are important in the non-perturbative regime, when $|L_\e| \approx l_P$, since then one cannot use the standard perturbative techniques.

We would like to add some final remarks. The PI measure (16) has been choosen such that the following conditions are satisfied

1) the corresponding path integral is finite for a finite triangulation,

2) the EA equation has a perturbative in $\hbar$ solution, such that the classical limit is the Regge action plus the corresponding SM action (this  requires that $(\log(\mu))'' < 0$ for $|L_\e| \to \infty$, see \cite{M}),
  
3) the non-smooth corrections in the perturbative effective action coming from the PI measure should be subleading to the smooth corrections, so that the leading quantum corrections are given by a QFT with a cutoff \cite{M1}.

There are other choices of the PI measure that satisfy the conditions 1), 2) and 3), but (\ref{plm}) is the most straigthforward choice. The diferent choices of the PI measure will lead to different physical predictions. In the case of the measure (\ref{plm}), the free parameters $(L_0, l_0, p)$ will be fixed by observations and experiments. For example,  by using the observed value of the cosmological constant one can fix $L_0$ to be of the order of $10^{-4}m$, see \cite{MVb, MVcc}.

Note that the measure (\ref{plm}) only affects the gravitational DOF, while the matter fields have the standard PI measure, consistent with the standard Quantum Mechanics \cite{M2}. 

Although the presented PFQG theory is finite and has the correct semiclassical limit, this does not mean that such a theory is the quantum gravity theory of our universe. It is only a candidate theory, like the superstring theory or CDT,  so that the PFQG theory has to be tested against the observations and the experiments. The physical predictions of a PFQG theory will depend on the choosen triangulation and on the parameters of the measure. 

Note that the dominant contribution in the semiclassical approximation of the PFQG effective action, when the edge lengths are microscopically small but larger than $l_P$, is given by a QFT on a smooth spacetime. This effective QFT is triangulation independent but it depends on the parameters of the PI measure, which can be then constrained by the values of the physical constants, like the cosmological constant.

In the PFQG case there is no need to sum over the spacetime triangulations, since a chosen triangulation is considered to be a physical property of the spacetime. Hence one can in principle discover the underlying triangulation of the spacetime by doing appropriate experiments. For example, the amplitude of a very high energy particle scattering (energies between $10^5$ GeV and $10^{19}$ GeV)\footnote{The Large Hadron Collider center of mass energy is $1,4\cdot 10^4$ GeV, while the Planck energy is $1,2\cdot 10^{19}$ GeV.} could in principle give the information about the spacetime triangulation via the triangulation dependent terms in the effective action, which are less supressed for higher energies (shorter distancies). This is analogous to the gamma ray scattering through a material, which can reveal the crystal lattice structure.

\bigskip
\bigskip
\bigskip
\noindent {\bf \large Acknowledgements}

Partially supported by by the Science Fund of the Republic of Serbia, grant 7745968, Quantum Gravity from Higher Gauge Theory 2021, QGHG-2021.


\end{document}